\documentclass[aps,prb,twocolumn,reprint,superscriptaddress]{revtex4-1}

\usepackage{graphicx}
\usepackage{amsmath}
\begin{document}

\title{Structural Simplicity as a Restraint on the Structure of Amorphous Silicon}

\author{Matthew J. Cliffe}
\email[]{mjc222@cam.ac.uk}
\affiliation{Department of Chemistry, Lensfield Road, University of Cambridge, CB2 1EW, UK}
\affiliation{Department of Chemistry, South Parks Road, University of Oxford, OX1 3QR, UK}

\author{Albert P. Bart\'{o}k}
\affiliation{Department of Engineering, Trumpington Street, University of Cambridge, CB2 1PZ, UK}

\author{Rachel N. Kerber}
\affiliation{Department of Chemistry, Lensfield Road, University of Cambridge, CB2 1EW, UK}

\author{Clare P. Grey}
\affiliation{Department of Chemistry, Lensfield Road, University of Cambridge, CB2 1EW, UK}

\author{G\'{a}bor Cs\'{a}nyi}
\affiliation{Department of Engineering, Trumpington Street, University of Cambridge, CB2 1PZ, UK}

\author{Andrew L. Goodwin}
\affiliation{Department of Chemistry, South Parks Road, University of Oxford, OX1 3QR, UK}

\date{\today}

\begin{abstract}
Understanding the structural origins of the properties of amorphous materials remains one of the most important challenges in structural science. In this study we demonstrate that local `structural simplicity', embodied by the degree to which atomic environments within a material are similar to each other, is powerful concept for rationalising the structure of canonical amorphous material amorphous silicon (\emph{a}-Si). We show, by restraining a reverse Monte Carlo refinement against pair distribution function (PDF) data to be simpler, that the simplest model consistent with the PDF is a continuous random network (CRN). A further effect of producing a simple model of \emph{a}-Si is the generation of a (pseudo)gap in the electronic density of states, suggesting that structural homogeneity drives electronic homogeneity. That this method produces models of \emph{a}-Si that approach the state-of-the-art without the need for chemically specific restraints (beyond the assumption of homogeneity) suggests that simplicity-based refinement approaches may allow experiment-driven structural modelling techniques to be developed for the wide variety of amorphous semiconductors with strong local order.

\end{abstract}

\maketitle

\section{Introduction}
Amorphous materials are the crucial components of many next-generation technologies, including the high capacity anode material silicon \cite{Key2011} and the porous carbons used as supercapacitors \cite{Forse2015a} used for electrochemical storage, but despite their scientific and technological importance, many questions remain about their structures. This is due to the challenges both in creating realistic atomistic models of amorphous materials and in interpreting these models to uncover their ordering principles. Although diffraction data from amorphous materials lack Bragg peaks, these data remain some of the key sources of information about the structures of amorphous materials \emph{via} the total scattering structure factor and its Fourier transform, the pair distribution function (PDF), which are well-defined even without long-range order \cite{Egami2003}. Indeed, for disordered and nanoscale crystalline materials, advances in characterisation techniques have made refinement of crystal structures using the PDF a routine part of the analytical toolbox for problems from pharmaceuticals \cite{Chen2014,Davis2013} to nanosized catalysts \cite{Du2012,Funnell2014}. These techniques do however rely on the presence of some degree of periodic average structural order as a restraint. For amorphous materials not only are these analytical techniques inapplicable, but the large number of atoms necessary for a representative sample makes the interpretation of the resultant model more difficult. These twin challenges represent a significant barrier to our understanding of non-crystalline materials \cite{Juhas2015}. 

\begin{figure}
\includegraphics{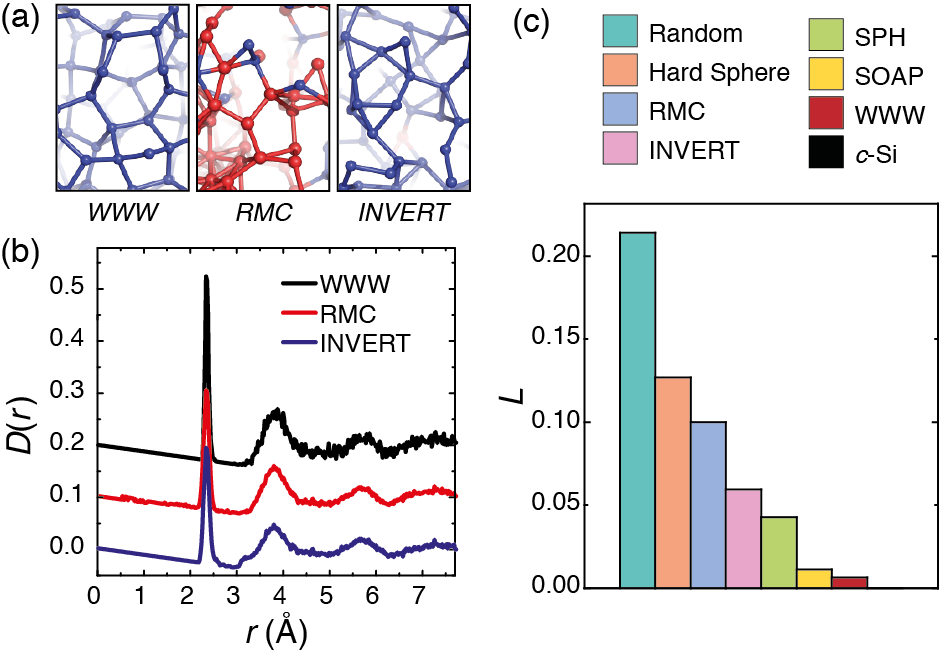}
 \caption{\label{fig:intro} 
 (a) Models of \emph{a}-Si with equivalent PDFs: the high-quality WWW model; a model produced by RMC fitting to data also using the INVERT PDF-variance restraint; and a model produced by RMC fitting to data with no additional restraints. Four-coordinate Si atoms are shown in blue, miscoordinated atoms are shown in red. (b) The calculated PDFs for these models are very similar (shown in the $D(r)$ normalisation \cite{Keen2001}). (c) $L$, our measure of simplicity shows that these data-derived models (RMC, INVERT) are more complex than the WWW model and less complex than the more disordered models (random, hard sphere). The two new models produced using RMC refinement with similarity restraints (the $L$ and SPH models) are closer in complexity to the WWW model.}
 \end{figure}

The reverse Monte Carlo (RMC) algorithm is one of the most popular methods for producing atomistic models from experimental data as it can produce large (thousands of atoms) supercells consistent with a given set of data (typically diffraction data) through iterated small random atomistic moves \cite{McGreevy1988,Playford2014}. The randomness inherent in the RMC algorithm causes the refined models to contain the maximum amount of disorder that is consistent with the experimental data. Therefore, because diffraction data only contain information on pairwise correlations, RMC refinement against them alone only produces appropriate structural models where the important interactions are also predominantly pairwise: for example noble gas liquids \cite{McGreevy1988} or metallic glasses \cite{Sheng2006}. In most functional materials higher-order terms make significant contributions to the energetics of the material, so refinement against just diffraction data will fail in the absence of long-range periodicity \cite{Evans1990,Welberry1994}. The paradigmatic example of this failure is \emph{a}-Si, where the presence of higher order correlation terms lead unconstrained RMC refinement against diffraction data to produce highly unphysical models which nevertheless reproduce the diffraction data to the same extent as physically-sensible models, as illustrated here by the WWW and RMC models \cite{Wooten1985,Gereben1994} [Fig.~\ref{fig:intro}(a,b)]. For amorphous semiconductors this failure is most starkly illustrated by the absence of an electronic band gap in models constrained only by the average pair correlations.

To understand the structural origins of the electronic properties of these materials we must therefore make use of information beyond the pair correlations, both in the generation of models and in their interpretation. Spectroscopic techniques, especially nuclear magnetic resonance measurements, can be exceptionally sensitive to these higher order correlations, but the structural information contained within these spectra is very often not transparently accessible. Thus, quantum chemical calculations, in particular density functional theory, are typically required to extract it. The expense of these calculations has meant most success has been found through using spectroscopic measurements to validate proposed models rather than to inform their creation \cite{Mayo2016}. Quantum chemical calculations can also be extremely valuable in their own right, as they intrinsically incorporate accurate information about the higher order interactions in materials. These calculations remain very computationally intensive for the large system sizes needed to accurately describe amorphous materials, although recent work has shown that an approach combining RMC refinement with \emph{ab initio} relaxation can overcome the configurational barriers to reorganisation that have limited the application of quantum chemical calculations thus far, resulting in much more realistic models of amorphous materials \cite{Pandey2016,Pandey2016a}. 

Alternatively, we can make use of assumptions about the anticipated geometric arrangements within the material to design empirical potentials, thus avoiding the computational expense of \emph{ab initio} calculations \cite{Soper2005,Opletal2013}. These can be very effective for cases where we already understand the nature of the interactions within a material, although they still often require more sophisticated algorithms to produce the highest quality models \cite{Barkema1996,Biswas2016}. For \emph{a}-Si, one of the simplest and most successful approaches has been the Wooten Winer Weaire (WWW) algorithm, which generates four-fold coordinated random networks by combining bond-switch moves with relaxation against a classical potential \cite{Wooten1985,Djordjevic1995,Barkema2000}. The WWW approach still provides the benchmark models of \emph{a}-Si, as judged by comparison with experimental diffraction, spectroscopic and electronic structural data \cite{Drabold2009}. Despite the practicality of empirical potentials, the assumptions inherent in using one potential rather than another can restrict both the generality of conclusions and the reliability of the results for new and poorly understood materials.

There is therefore still a need for methods that are able to introduce physical reasonableness without relying on detailed and expensive quantum chemical calculations. The characteristic failing of data-driven model building approaches has been that their stochastic nature leads to unphysical structural complexity in the resultant models \cite{Cliffe2010}. Modifying the RMC algorithm to favour simpler solutions should therefore produce more realistic models. Indeed, biasing the refinement to such that the variance in atomic PDFs is also minimised (the INVERT approach which embodies the assumption that all atoms should have similar pair correlations) did allow RMC to produce models of \emph{a}-Si and \emph{a}-SiO$_2$ with improved structural properties, although these configurations were still lacking in some key electronic properties (\emph{e.g.} absence of any band gap) [Fig.~\ref{fig:intro}] \cite{Cliffe2010}.

In this study we explore the role of structural simplicity in \emph{a}-Si, a canonical example of an amorphous semiconductor. First we show that the degree of structural variance of local environments in a model, measured by the recently developed smooth overlap of atomic positions (SOAP) descriptor, can quantify the structural simplicity in models of \emph{a}-Si \cite{Bartok2013,De2016}. We then go on to show that using this new measure of simplicity as a restraint on reverse Monte Carlo refinement against ideal PDF data does drive RMC to produce much simpler models, and that these simpler models are more physical as assessed both by structural correlation functions and electronic structures. These simpler models are of sufficient quality that DFT minimisation of the resultant configurations yields models that qualitatively reproduce both the structural and electronic features of the highest-quality models of \emph{a}-Si. We finally go on to show that the reverse relationship also holds: that making a model of \emph{a}-Si more physical also tends to make it structurally simpler.

\section{The Smooth Overlap of Atomic Positions descriptor}
We begin by briefly introducing the SOAP descriptor \cite{Bartok2013,De2016}. Perhaps the most direct way of comparing the similarity of two structures is to superimpose one on the other and examine the degree of spatial overlap between the two. However, for an amorphous material there is no meaningful orientational or translational frame of reference. This approach has been developed to allow the evaluation of the degree of spatial overlap between two environments without needing to specify the orientational relationship between them. The SOAP degree of similarity between the local environments of atoms $i$ and $j$, $k_{ij}$, is defined as the integral over positions, $\mathbf{r}$, and all rotations, $\hat{R}$, of the product of the two local atom densities $\rho_i$ and $\rho_j$ :

 \begin{equation}
 \label{eqn:SOAP_dist_krr}
	k_{ij}\,=\, \int \! \mathrm{d}\hat{R} \left | \int \mathrm{d}\mathbf{r} \rho_i(\mathbf{r}) \rho_j(\hat{R}\:\mathbf{r}) \right|^2.
 \end{equation}
To ensure that $k_{ij}$ is a smooth function, the atomic density is convolved with Gaussian broadening function with a width $\alpha$, and the local nature of this density is ensured by applying smooth radial cut-off function, $f_\textrm{cut}(r_{ij})$. The atomic density $\rho_i(\mathbf{r})$ is thus 
\begin{equation}
\rho_i(\mathbf{r})\,=\, \sum_j\, f_\textrm{cut}(\left| \mathbf{r}-\mathbf{r}_{ij} \right|)~\mathrm{exp}(-\alpha \left| \mathbf{r}-\mathbf{r}_{ij} \right| ^2).
 \end{equation}
It is often helpful for many applications to normalise $k_{ij}$ such that the self-similarity of any environment is one, which yields the metric $K_{ij}$:
 \begin{equation}
 \label{eqn:SOAP_dist_Krr}
	K_{ij}\,=\, \frac{k_{ij}}{\sqrt{k_{ii}k_{jj}}} .
 \end{equation}
$K_{ij}$ can be evaluated by expanding the angular dependence of $k_{ij}$ as a series of spherical harmonics and expanding the radial component using a series of orthogonal radial basis functions. The derivation and exact form of this expansion can be found in Ref.~\citenum{Bartok2013}. In the present work we have used a Gaussian smoothing $\alpha = 0.5${\,\AA}, a radial cut-off of 3.0{\,\AA}, which places the smooth cut-off between the first two peaks in the PDF, and made use of the derived metric $L_{ij}\,=\,-\mathrm{log}\,K_{ij}$ to reduce the effect of outlying values. Although in this paper we have made use of this metric to study elemental silicon, it is not limited to monoelemental systems. One direct approach would be to consider partial descriptors for individual atomic pairs; for example in SiO$_2$ one could consider the Si--Si, O--O, Si--O and O--Si descriptors separately. A more sophisticated application of this metric would be making using of a composite `alchemical' descriptor, for which an additional alchemical similarity metric $\kappa_{\alpha\beta}$ is defined for each pair of elements $\alpha$ and $\beta$. This alchemical similarity measure has already been successfully applied to cluster a series of molecules according to their chemical and structural similarities \cite{De2016}.

The validity of our self-similarity metric was checked by calculating $L \,=\, \sum_{ij} L_{ij}$ for six candidate models of \emph{a}-Si which possess varying degrees of order (listed from least to most ordered)\footnote{See Supplementary Material at URL for the coordinates of both these reference configurations and those created during this study, along with further details of all described configurations}:
\begin{description}
\item[Random] A random configuration with no other restraints.
\item[Hard sphere] A random configuration generated with the restraint that no atom be placed within 2.2\,{\AA} of another.
\item[RMC] A random configuration generated through RMC refinement against PDF data with no other restraints.
\item[INVERT] A random configuration generated through RMC refinement against PDF data with the INVERT PDF variance restraint applied \cite{Cliffe2010}.
\item[WWW] A configuration generated using the WWW algorithm \cite{Barkema2000}.
\item[\emph{c}-Si] Crystalline diamondoid silicon.
\end{description}
This $L$ value measures the variation between local atomics environments within the configuration: a large value of $L$ results from a high diversity of environments (low simplicity), and a small value from a low diversity of environments (high simplicity), and it produced the same ranking of simplicity, distinguishing between the three configurations with equivalent PDFs [Fig.~\ref{fig:intro}(c)]. 

\section{RMC Refinement}
On this basis, we proceeded to explore whether this metric could be used as a restraint on RMC refinement against pair distribution function data for \emph{a}-Si calculated from a high quality WWW generated model \cite{Barkema2000}. A starting model of 512 atoms randomly distributed throughout a 21.7\,{\AA} cubic box with periodic boundary conditions was used, and then fitted to the pair distribution function by optimising the following objective function using simulated annealing and small individual atomic moves:

\begin{multline}
\chi = \frac{w_\mathrm{PDF}}{N} \sum_j\sum_r \frac{[g_j(r) - g_\mathrm{expt}(r)]}{r^2} \\ + \frac{w_L}{N} \sum_{ij}\,L_{ij},
\end{multline}

where $N$ is the total number of atoms, $g_j(r)$ is the individual atomic radial distribution function, $g_\mathrm{expt}$ is the experimental radial distribution function and $w_\mathrm{PDF}$ and $w_L$ are weightings for the PDF data and self-similarity restraint $L$, respectively.

\begin{figure}
 \centering
 \includegraphics[scale=0.35]{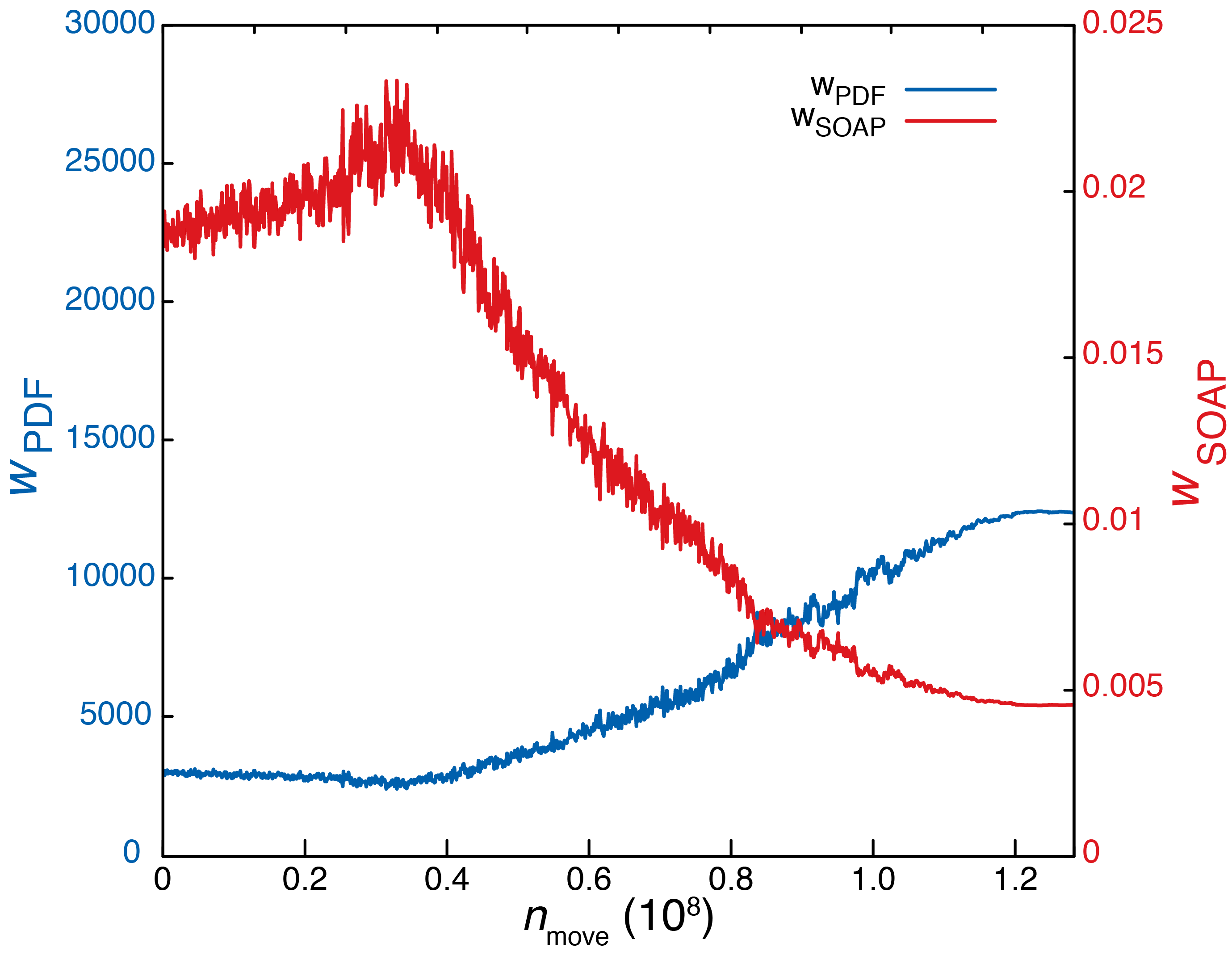}
  \caption[Evolution of weights]{Evolution of the weighting scheme for the joint simplicity and PDF refinement using the dynamic weighting scheme throughout the refinement as a function of proposed moves.}
 \label{fig:weight}
\end{figure}

As well as refining against the self-similarity measure $L$, we carried refinements against the spherical harmonics measures of similarity described in Ref.~\citenum{Cliffe2013}. In order to produce significant improvements over the models refined against PDF+INVERT, it was necessary to also include both the spherical harmonics variance $Q_l$ and the measure of local symmetry $S$ as restraints. We encountered a number of difficulties during these refinements which are well known for constrained RMC refinements: first, the need to choose weighting factors ($w$) and second, the low acceptance rate for proposed moves at low temperatures. Choosing appropriate weights proved very important, not only for the multi-restraint spherical harmonics+INVERT+symmetry+PDF composite refinement, where there are four independent weighting factors, but also for the simpler refinement against just PDF+$L$. We found that a `design of experiments' approach was reasonably effective in allowing us to tune the relative weights for the composite refinement to produce a good fit to these metrics.
We found that the relative importance of the contributions of $L$ and PDF data to the objective function changed gradually throughout the refinement. To ensure that both contributed approximately equally to the objective function throughout the entirety of the refinement, \emph{i.e.} that $w_\mathrm{PDF} \chi_\mathrm{PDF} \approx w_L \chi_L$, after every temperature $w_\textrm{PDF}$ and $w_L$ were adjusted by a scaling factor:
\begin{equation}
A = \left( \cfrac{w_\mathrm{PDF} \chi_\mathrm{PDF}} { w_L \chi_L}\right)^{0.25},
\end{equation}
yield a new PDF weight $w'_\textrm{PDF}\,=\,\frac{w_\textrm{PDF}}{A}$ and a new self-similarity weight $w'_L\,=\,A\, w_L$. The evolution of the weights throughout the refinement can be seen in Fig.~\ref{fig:weight}.
This adaptive weighting scheme also helped ameliorate the low acceptance rate that is characteristic of these constrained RMC refinements, though the refinements still required a large number of moves to converge \cite{Cliffe2013a}. Although in both cases we were able to obtain good fits to both the data and restraints, the $L$+PDF model was of higher quality and also conceptually simpler than the spherical harmonics+INVERT+symmetry+PDF composite refinement, and so for the remainder of this article we will focus on that model. It is important to note that in this refinement we made no assumptions about the expected local environments, \emph{e.g.} tetrahedral geometry or four-fold coordination, beyond that they should be similar to each other (an assumption appropriate for \emph{a}-Si).

\begin{figure}

\includegraphics{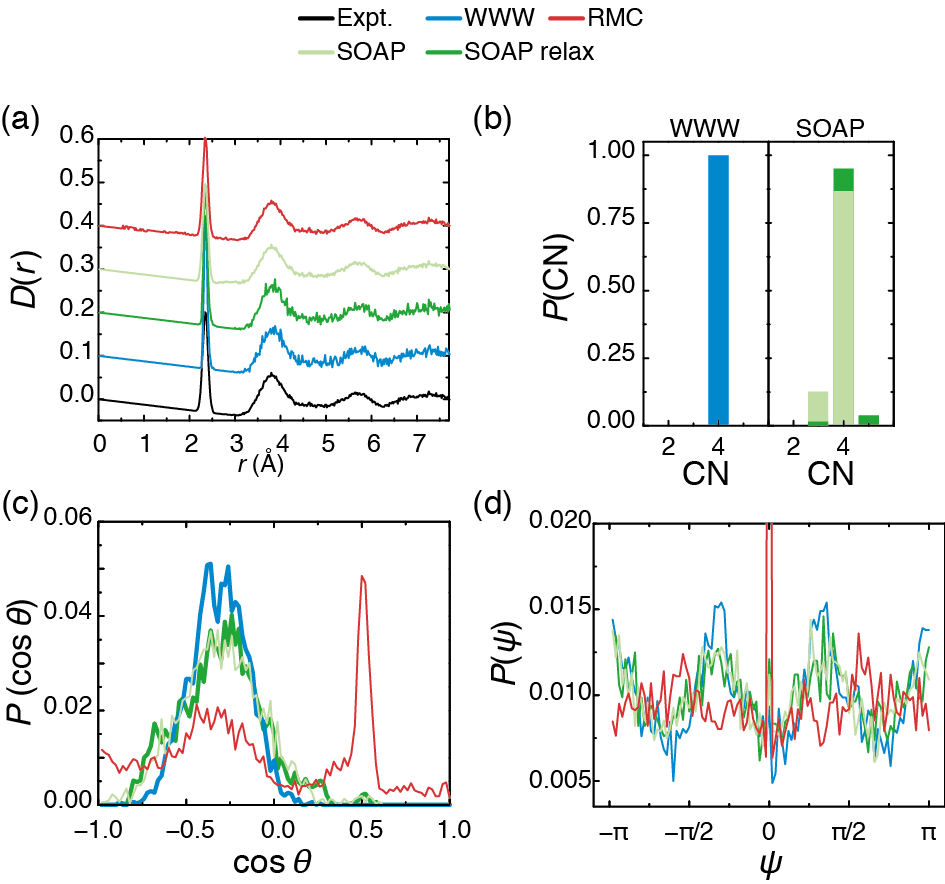}
\caption{\label{fig:model_geom} Key geometric correlation functions for the different models of \emph{a}-Si: WWW, the refined $L$ configuration, the $L$ model after relaxation, and an RMC refined configuration with no other restraints. (a) Pair distribution functions ($D(r)$ normalisation) including comparison to the referenced WWW-derived model, (b) the coordination number distribution $P(\mathrm{CN})$, (c) the bond angle distribution $P(\mathrm{cos}\,\theta)$ and (d) the dihedral angle distribution $P(\psi)$. The dihedral angle distribution for the RMC model has been truncated as it reaches a peak of 0.077 at $\psi\,= \,0\,^\circ$.}
 \end{figure}

Examination of this model refined against PDF and $L$ revealed that, in addition to fitting the PDF well, it reproduced the general features of the higher-order correlation functions [Fig. \ref{fig:model_geom}]. This is clearest in the bond angle distribution, which has no peak at $\mathrm{cos}(\theta)=0.5$, indicating that unlike previous data-driven RMC models, there are very few unphysical Si$_3$ triangles [Fig. \ref{fig:model_geom}(c)] \cite{Kugler1993}. The dihedral angle ($\psi$) distribution of this model also shows the threefold symmetry indicative of tetrahedral coordination, and additionally confirms the elimination of Si$_3$ triangles (which produce a sharp peak at $\psi\,= \,0\,^\circ$) [Fig. \ref{fig:model_geom}(d)]. 

 \begin{figure}
\includegraphics{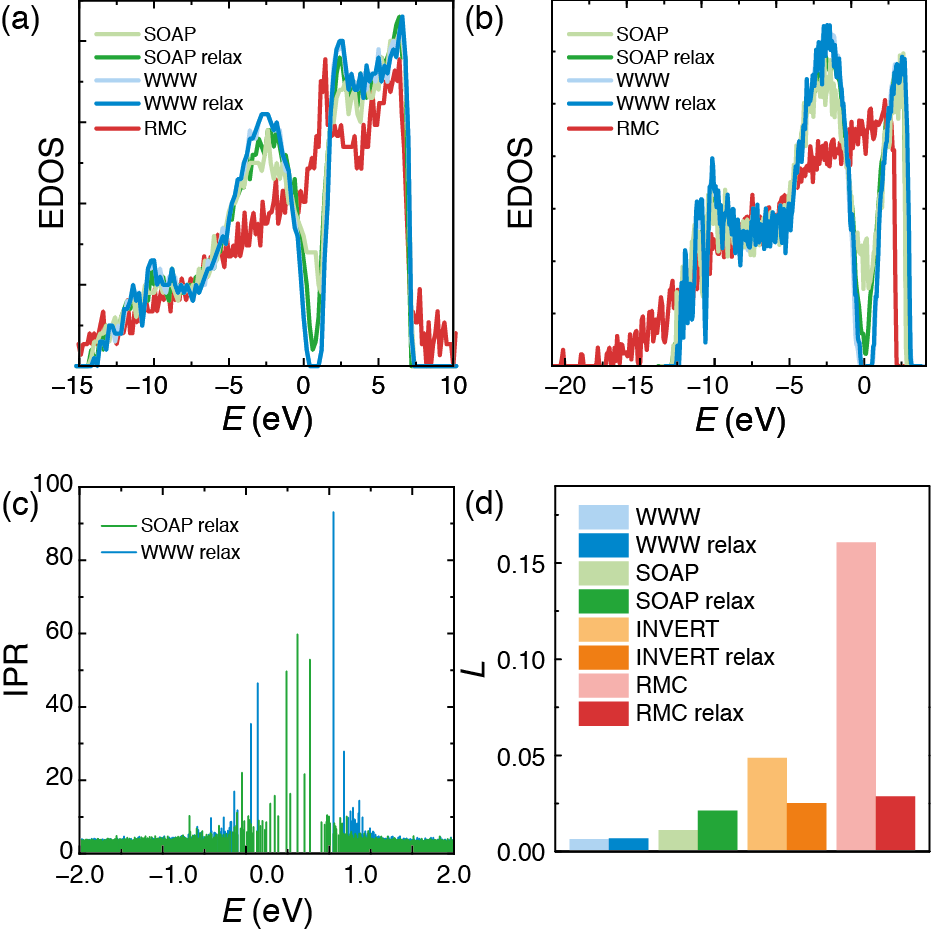}
 \caption{\label{fig:elec} (a) The calculated tight-binding (TB) electronic density of states (EDOS) for the $L$ and WWW models before and after \emph{ab initio} optimisation. The density of states for the unrestrained RMC refined configuration is shown for comparison. (b) The EDOS calculated using the MBJ functional, which reproduces the same form as the TB model ($E_F=0$ for the WWW optimised model). (c) The inverse participation ratio calculated from the TB structure for the optimised configurations and WWW configurations showing the localisation of near- and in-gap states. (d) Calculated SOAP measure $L$ for configurations before and after optimisation show that an increase in structural quality tends to also increase the structural similarity.}
 \end{figure}

\section{Electronic structure calculations}
Calculation of the electronic structures of these models using a standard tight-binding Goodwin-Skinner-Pettifor) Hamiltonian showed the dramatic improvement in the quality of the simplicity-refined model over the unconstrained RMC model [Fig.~\ref{fig:elec}(a)] \cite{Goodwin1989}. Unlike the RMC model, for which the EDOS is comparatively featureless, the EDOS of our new model qualitatively replicates the EDOS of the WWW model. However, as might be expected from the large number of dangling-bond coordination defects present [Fig. \ref{fig:model_geom}(b)], the SOAP model still possesses a significant density of gap states.

On this basis we decided to examine the electronic structure of this simplicity-refined model more closely using quantum mechanical calculations, both to validate our tight-binding calculations and to gain more detailed insight into their structures. We carried out \emph{ab initio} calculations using the projected augmented wave (PAW) method implemented in the Vienna \emph{Ab initio} Simulation Package (VASP) with a cut-off energy of 500 eV, evaluated at the $\Gamma$-point \cite{Kresse1993,Kresse1996,Kresse1999,Blochl1994}. We made use of the MBJ meta-GGA functional as it is able to capture accurately the electronic properties of semiconductors \cite{Tran2009}. These calculations confirmed the validity of our tight binding calculations. They also demonstrated the improvement in the energetics of our model compared to previous data-derived models: it has an energy of 0.24$\,$eV atom$^{-1}$ above WWW, whereas the RMC model has an energy of 6.20 $\,$eV atom$^{-1}$. 

To explore the reverse question, \emph{i.e.} whether improving the realism of a model also increases the structural simplicity, we optimised a number of configurations of \emph{a}-Si (RMC, INVERT, our new model and WWW) using the PBE functional \cite{Perdew1996,Perdew1997}. As expected, optimisation of the WWW model left it essentially unchanged (median atomic displacement $d\,=\,$0.024\,{\AA}). All three data-derived configurations underwent significant structural rearrangements: for our model, $d\,=\,$0.35\,{\AA}, for INVERT. $d\,=\,$0.97\,{\AA}, for RMC, $d\,=\,$1.35\,{\AA}. Remarkably, all three configurations converged to similar final structures (energies within 0.04\,eV atom$^{-1}$), with good electronic and structural properties, although the INVERT and RMC derived models required significantly more computer time to converge and retained slightly higher energies and concentrations of coordination defects. The relaxation also led to a slight degradation of the quality of fit to the PDF data, which was more severe for the INVERT and RMC models. Comparison of the optimised configuration with that derived purely from RMC refinement showed that optimisation had eliminated the overwhelming majority of the dangling-bond under-coordination defects, which in turn led to a reduction in the density of gap states [Fig. \ref{fig:elec}(a,b)]. Examination of the inverse participation ratio for the peri-gap states showed that the remaining gap states were highly localised [Fig. \ref{fig:elec}(c)]. These optimised models also all showed low values of $L$, confirming the close link between simplicity and physicality, although there was a small increase in $L$ for our model and WWW models due to the interplay between the energy and simplicity measures [Fig. \ref{fig:elec}(d)].

\section{Conclusions}
In this study we have shown that a general criterion of structural simplicity is a powerful restraint on the range of feasible structures. We have shown that when this restraint is applied to a canonical example of an amorphous material, \emph{a}-Si, it is able to guide the refinement to a primarily tetrahedral random network from diffraction data alone, without the need for any assumptions about the expected local geometry, showing that the simplest model of \emph{a}-Si consistent with the pair correlations is a CRN possessing an electronic pseudogap. \emph{Ab initio} optimisation is able to eliminate the vast majority of the remaining structural defects, producing models that are comparable to the state-of-the-art, both structurally and electronically, again without the need for system specific assumptions. This result suggests that parsimony may provide sufficient restraint for useful structural refinement against diffraction data for amorphous materials where the assumption of local homogeneity is expected to hold, \emph{e.g.} \emph{a}-P \cite{Zaug2008} or the amorphous transparent conducting oxides \cite{Walsh2009}.

The important role of electronic homogeneity in producing structural homogeneity has recently been demonstrated \cite{Prasai2015}, and our findings finally establish the converse relationship: that structural homogeneity tends to produce electronic homogeneity, at least for \emph{a}-Si. Previous work has shown \emph{a}-Si may in fact show hyperuniformity, a long-range non-periodic order, and that the degree of hyperuniformity in a model of \emph{a}-Si is closely linked to how well-relaxed the model is (\emph{i.e.} how closely the environment conforms to tetrahedrality)\cite{Torquato2003,Hejna2013,Xie2013,Torquato2016}. It has also been demonstrated that for two dimensional systems, the degree of local order is closely linked to its hyperuniformity, and this in turn is linked to the existence of a photonic band gap \cite{Florescu2009,Froufe-Perez2016}, although the causal relationship between hyperuniformity and the photonic band gap remains unclear \cite{Imagawa2017,Sellers2017}. The relationship established here between the homogeneity of local environments and the electronic band structure thus further emphasises the importance of non-periodic order in amorphous materials for their reciprocal space properties. The success of simplicity, as parametrised by the SOAP self-similarity, $L$, as a restraint on models of disordered structures also provokes questions about how to apply formal definitions of simplicity (and its converse, complexity) in these materials \cite{Cartwright2012,Krivovichev2014,Crutchfield2011}.

\begin{acknowledgments}
We gratefully acknowledge financial support, from Sidney Sussex College, Cambridge to M.J.C.; from EPSRC to C.P.G. and R.P.K. under Grant No. EP/K030132/1; from EPSRC (EP/G004528/2) and ERC (Grant Ref: 279705) to M.J.C and A.L.G.. A.P.B. was supported by a Leverhulme Early Career Fellowship with joint funding from the Isaac Newton Trust. Via our membership of the UK's HEC Materials Chemistry Consortium, which is funded by the EPSRC (EP/L000202), this work used the ARCHER UK National Supercomputing Service (http://archer.ac.uk). M.J.C. would like to thank J.A.M. Paddison for useful discussions.
\end{acknowledgments}

\bibliography{asi_SOAP_16}

\end{document}